\documentclass[sigconf]{acmart}

\usepackage{enumitem}
\usepackage{subfigure}
\usepackage{multirow}
\newcommand{\eat}[1]{}
\AtBeginDocument{%
  }

\copyrightyear{2023}
\acmYear{2023}
\setcopyright{acmlicensed}\acmConference[SSDBM 2023]{35th International Conference on Scientific and Statistical Database Management}{July 10--12, 2023}{Los Angeles, CA, USA}
\acmBooktitle{35th International Conference on Scientific and Statistical Database Management (SSDBM 2023), July 10--12, 2023, Los Angeles, CA, USA}
\acmPrice{15.00}
\acmDOI{10.1145/3603719.3603734}
\acmISBN{979-8-4007-0746-9/23/07}

\begin{document}

\title{Privacy-Preserving Redaction of Diagnosis Data  through~Source~Code~Analysis}

\newcommand{\tsc}[1]{\textsuperscript{#1}} %

\author{Lixi Zhou}
\authornote{This work was done when Lixi Zhou and Lei Yu work in IBM.}
\affiliation{
  \institution{Arizona State University}
  \country{United States}
}

\author{Lei Yu}
\authornotemark[1]
\affiliation{
  \institution{Rensselaer Polytechnic Institute}
  \country{United States}
}

\author{Jia Zou}
\affiliation{
  \institution{Arizona State University}
  \country{United States}
}

\author{Hong Min}

\affiliation{
  \institution{IBM T. J. Watson Research Center}
  \country{United States}
}

\begin{abstract}
  Protecting sensitive information in diagnostic data such as logs, is a critical concern in the industrial software diagnosis and debugging process. While there are many tools developed to automatically redact the logs for identifying and removing sensitive information, they have  severe limitations which can cause either over redaction and loss of critical diagnostic information (false positives), or disclosure of sensitive information (false negatives), or both. To address the problem, in this paper, we argue for a source code analysis approach for log redaction. To identify a log message containing sensitive information, our method locates the corresponding log statement in the source code with logger code augmentation, and checks if the log statement outputs data from sensitive sources by using the data flow graph built from the source code. Appropriate redaction rules are further applied depending on the sensitiveness of the data sources to preserve the privacy information in the logs. We conducted experimental evaluation and comparison with other popular baselines. The results demonstrate that our approach can significantly improve the detection precision of the sensitive information and reduce both false positives and negatives.
\end{abstract}

\eat{
\begin{CCSXML}
<ccs2012>
 <concept>
  <concept_id>10010520.10010553.10010562</concept_id>
  <concept_desc>Computer systems organization~Embedded systems</concept_desc>
  <concept_significance>500</concept_significance>
 </concept>
 <concept>
  <concept_id>10010520.10010575.10010755</concept_id>
  <concept_desc>Computer systems organization~Redundancy</concept_desc>
  <concept_significance>300</concept_significance>
 </concept>
 <concept>
  <concept_id>10010520.10010553.10010554</concept_id>
  <concept_desc>Computer systems organization~Robotics</concept_desc>
  <concept_significance>100</concept_significance>
 </concept>
 <concept>
  <concept_id>10003033.10003083.10003095</concept_id>
  <concept_desc>Networks~Network reliability</concept_desc>
  <concept_significance>100</concept_significance>
 </concept>
</ccs2012>
\end{CCSXML}

\ccsdesc[500]{Computer systems organization~Embedded systems}
\ccsdesc[300]{Computer systems organization~Redundancy}
\ccsdesc{Computer systems organization~Robotics}
\ccsdesc[100]{Networks~Network reliability}
}

\maketitle

\section{Introduction}\label{sec:introduction}

Diagnostics data (e.g., logs and traces) is generated by a wide variety of systems and devices, including operating systems, applications, devices, and vehicles. It is incredibly valuable for improving system performance, troubleshooting, and reducing maintenance cost\cite{sanchez2016c}. For example, error logs and crash reports can help developers and technicians identify software bugs and fix system issues, and \nobreak network diagnostic data can reveal the issues with connectivity or bandwidth. Therefore, it is often shared between different parties, such as software developers, device manufacturers, cloud service providers, etc., for those purposes. 

However, diagnostics data can raise serious privacy concerns since it may contain personally identifiable information (PII) or sensitive data. PII is any information that can be used to identify an individual, such as name, social security number and bank account number. Sensitive data could be personal health information (PHI), financial information and credential data. On the other hand, today's regulatory compliances such as Health Insurance Portability and Accountability Act (HIPAA) \cite{annas2003hipaa}, Payment Card Industry Data Security Standard (PCI-DSS) \cite{virtue2009payment} and General Data Protection Regulation (GDPR) \cite{regulation2018general}, place strict limitations on how this information can be collected and shared. Thus, diagnostics data privacy has become a severe obstacle in industrial collaborative software diagnosis and debugging processes, and it is critical for organizations to take steps to protect sensitive information in diagnostics data and prevent inadvertent disclosure.

A common approach to protecting diagnostics data privacy is log redaction. This technique involves removing or obfuscating sensitive information from diagnostic logs before they are shared. Existing log redaction mechanisms typically use two strategies: (1) Rule-based approach~\cite{mouza2010towards, park2011experimental, chow2008detecting, amazon-macie, google-dlp} that detects private information based on a set of pre-defined string patterns matching or regular expressions; (2) Machine learning-based approach \cite{kaul2021knowledge, cumby2011machine, jones2007know, microsoft-presidio} that trains a model to automatically identify and redact sensitive information from logs. However, there are several limitations to these strategies that hinder their effectiveness:
\begin{itemize}[leftmargin=*]
\item Rule-based approach relies on predefined rules and thus is not effective and flexible for identifying unseen or modified types or patterns of sensitive information.
\item Machine learning based approach requires a large amount of training data to be effective and the labeling is labor-intensive and time-consuming \cite{he2020loghub}.
\item Both approaches can produce false positives when non-sensitive information is mistakenly identified as sensitive, and false negatives when sensitive information is not identified. This can cause either over redaction and loss of important diagnostic information, or disclosure of sensitive information.   

\end{itemize}

To address these limitations, in this work, we argue for a novel source code based approach to log redaction by analyzing and tracking the data flow for each logging/tracing statement. It has been observed that open source software constitutes 70-90\% of any given piece of modern software solutions~\cite{open-source-summary}, but the source code is highly underutilized for diagnostics data processing.
Our proposed methodology leverages it to identify the data sources of a log message and apply the corresponding redaction rule based on its sensitiveness. Our approach extracts the structured data flow information from the source code and matches such information with the logging/tracing output, which leads to higher precision and recall compared to existing rule-based or learning-based approaches. 
If the information shares a similar pattern, it is very likely that the non-sensitive information will be redacted, which is considered a false positive match. For example, not all email addresses should be redacted, and those email addresses (e.g., belong to suspect adversaries) which are related to the diagnosis/auditing purposes should not be redacted. The source code analysis can help to classify the sensitiveness of the data source and that information can be leveraged to reduce the false positives and false negatives.

\subsection{Challenges}
Despite the opportunities, our source code analysis approach needs to address two challenging problems.

\noindent
\textbf{$\bullet$ Link log statements to sensitive data sources}. A log statement usually prints program variables in a formatted string. Those variables may contain results directly or derived from sensitive sources at run-time. It is not practical and efficient to dynamically track information flow at run-time. Since log redaction post-processes logs, it is necessary to pre-build and store the linkage information between log statement variables and data sources and support efficient on-demand queries during post-processing. To address this problem, we apply static analysis techniques and build a data flow graph (DFG) from the source code. Although data flow graph has been already widely used for sensitive information tracking~\cite{chao2023flow}, our approach aims to build a succinct data flow graph tailored for diagnostics data to facilitate efficient redaction.  

\noindent
\textbf{$\bullet$ Map log message to log statement}. To use the linkage between log statements and sensitive data sources for log redaction,  our approach requires efficiently mapping a log message to its corresponding log statement in the source code. A general approach is to mine the log templates that consist of constant keywords in the print statements from log messages and match a log message's template to the log statement code. However, this approach incurs additional post-processing overhead for template mining. Given that much open-source software uses standard logger packages, in this work we explore an alternative approach for simplicity that exploits logger configuration to directly output the location of log statement for a log message.

\eat{
}

\subsection{Our Contributions: A Novel Redaction Framework based on Source Code Analysis}

\begin{figure*}[!h] 
\centering
\includegraphics[width=0.9\textwidth]{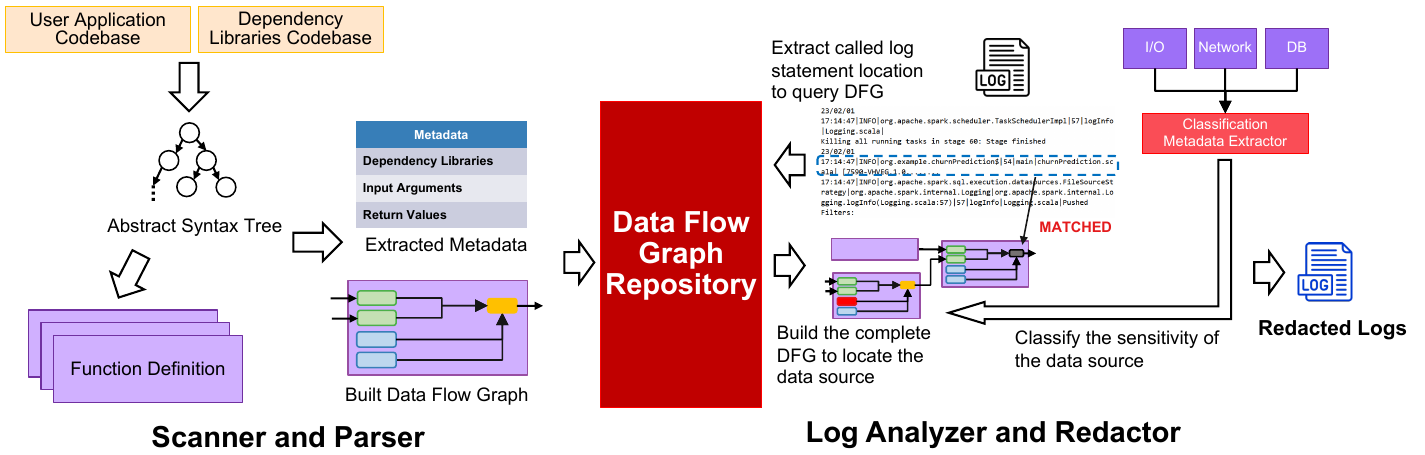}
\vspace{-5pt}
\caption{\label{fig:overview} \small
Overview of proposed methodology
}
\vspace{-5pt}
\end{figure*}

To address these challenges, we proposed a novel log redaction framework based on source code analysis to effectively identify and redact the sensitive information from the diagnostics data, which is illustrated in Figure ~\ref{fig:overview}. The framework consists of three modules: Scanner/Parser, Data Flow Graph Repository, and Log Analyzer/Redactor.

\vspace{5pt}
\noindent
\textbf{(1) Scanner and Parser} This module builds a data flow graph from the source code to track the data source for each log message. It takes users' application source code, and the source codes of the linked third party libraries as input and then parse the source codes at the function level to build the data flow graph. This process consists of two steps:

\underline{$\bullet$ Step 1. Extracting Function Information} This step parses each source file into a collection of the function definitions. At the same time, the function name, arguments list, arguments datatypes, and return variables, return variable tables will be collected as metadata for each function definition. To facilitate the execution of queries over the data flow graph at the later stages, the module further creates a unique identifier for each function, which is used for fast indexing.

\underline{$\bullet$ Step 2. Building Data Flow Graph (DFG)}. This step takes the source code of each function definition, which is the output of the first step, as the input to build an abstract syntax tree (AST). Then a data flow graph is built on top of the AST. Each node represents a variable or statement, and each directed edge represents the actual data that is produced by the source node and consumed by the destination node. For each data flow graph, the flow will start with the arguments that are passed to this function and end with the return value node or the last statement node if there is no return value for the given function.

\vspace{5pt}
\noindent
\textbf{(2) Data Flow Graph Repository} which is proposed to effectively store and manage the data flow graphs, which are built by the first module. It stores the data flow graph (DFG) for each function as a key-value pair format, where key is its unique identifier, and the value is a pointer to the DFG. Before the DFG is stored in the repository, optimization techniques such as pruning will be applied to the data flow graph to  reduce the complexity of the DFG to further save the storage overhead and improve the performance of back-tracing at the log redaction stage.

\vspace{5pt}
\noindent
\textbf{(3) Log Analyzer and Redactor} takes the metadata of the data sources, the data flow graph, and the original logs as inputs; and outputs redacted logs. We assume the data sources are structured and sensitive attributes are identified and labeled by domain experts. To effectively apply appropriate reduction rules to each log message, it is important to identify the sensitiveness (or privacy level) of its data source. This is accomplished by analyzing each log message and locating its related log statement in the source file. Then, a full data flow path is identified by back-tracing the DFG from the log statement to its source nodes. While traversing the DFG for each log message, each subgraph in the DFG is retrieved based on the unique identifier and connected via the metadata, argument lists, and return value, which are collected from the source code parsing module. Once the data source is located, the corresponding redaction rules will be applied to the log message based on the sensitiveness of its data source. This process will be conducted for each log message iteratively. In the end, the module outputs the final redacted log.

\section{Implementation Details}\label{sec:method}

\subsection{Target Scenario and System Overview}
In our target scenario, software or a pipeline to be diagnosed will consume/access a collection of data sources, with each data source being structured and private/sensitive attributes being identified and annotated by domain experts. Then the software will generate diagnosis data such as logs and traces, which may contain private/sensitive information from the data sources. The goal of the proposed framework is to detect and redact log messages that contain private information.

Using a source code analysis approach, we are able to identify the data sources (e.g., attributes) that are linked to each log message and then we redact the log message, if the data sources contain sensitive information.

As aforementioned, we propose a novel diagnostic data redaction framework based on source code analysis. The framework will detect sensitive information from the diagnostic data by linking each log message to its data sources using a data flow graph extracted from the source code. To achieve this, each source file of the users' application (i.e., the software/pipeline that generates the diagnostic data)  will be scanned and parsed into a collection of data flow subgraphs at the function level. Each subgraph can be used to track how the data is transferred within a function. 

Once a data flow subgraph is built, it will be stored in the data flow graph repository. An indexing that maps each function's unique identifier to its data flow subgraph is also constructed. 

Then, the log redactor takes the original diagnostic data, the user annotated sensitive data source information, the source code, and the data flow graph repository as input. It first analyzes each log message to locate the position of the log statement that produces the message, in the source code. 

At the redaction time, it needs to query the data flow graphs to identify all invoked functions from the data flow graph repository. Afterwards, a complete data flow graph is assembled and the data source nodes of the current log message are also located by traversing the full data flow graph. The detected data sources are cross-checked with users' sensitiveness annotation of the data sources to determine whether the redaction is needed and which redaction rule should be applied. 

\subsection{Building Data Flow Graph}

\begin{figure}
    \centering
    \subfigure[Metadata]{\includegraphics[height=1.2in]{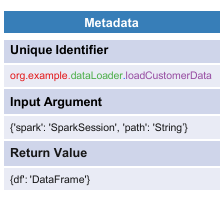}\label{fig:scan-and-parse-a}} 
    \subfigure[Data Flow Graph]{\includegraphics[height=1.2in]{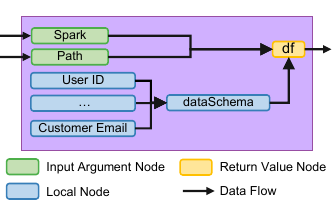}\label{fig:scan-and-parse-b}} 
    \vspace{-5pt}
    \caption{(a) Metadata extracted for each function by traversing its AST of the source code(b) The data flow graph built on its corresponding function definition
    }
    \label{fig:scan-and-parse}
    \vspace{-10pt}
\end{figure}

Data flow graph (DFG) is used to track the movement of the data across functions and variables, as illustrated in Figure~\ref{fig:scan-and-parse}. Given a log message that is produced by a logging statement, by traversing the data flow graph, the system can locate the attributes of the data sources that flow to the logging statement in question.  Then, based on the sensitiveness annotated for each data source attribute, the system obtains the sensitiveness of the log message. 

However, building a large-scale DFG for each complicated application may suffer from several pain points: (1) building a single DFG for the entire code base could be time-consuming,  and it is hard to maintain the DFG, if the source codes get frequently updated. (2) The time complexity required for traversing the DFG increases with the size of the DFG, which slows down the detection and redaction process. (3) The graphs associated with the shared libraries and the frequently invoked functions will be duplicated multiple times, which brings additional storage overhead. 

To address these issues, we decided to parse each source code file into a collection of function definitions and build a relatively small DFG for each function, and store it in the data flow graph repository for being queried at the redaction time. 

\begin{figure}[t]
    \centering
    \subfigure[Metadata to be extracted at parsing stage]{\includegraphics[width=3in]{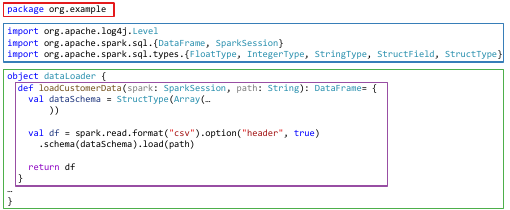}\hspace*{0.15in}\label{fig:source_code_parse-a}} 
    \subfigure[Abstract syntax tree of the source code]{\includegraphics[width=3in]{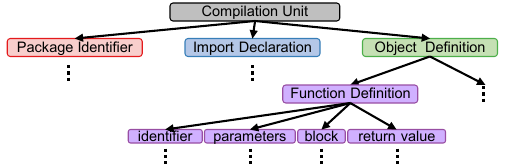}\label{fig:source_code_parse-b}} 
    \vspace{-5pt}
    \caption{(a) The metadata of each function needs to be extracted after parsing the source code file. (b) After converting the source code into an abstract syntax tree (AST), the corresponding metadata is extracted by traversing the tree.
    }
    \label{fig:source_code_parse}
    \vspace{-5pt}
\end{figure}

To achieve this, we track the data flow by analyzing the source code and parsing it into an abstract syntax tree (AST) \footnote{We use the tree-sitter as the parser to parse the source code into AST. https://tree-sitter.github.io/tree-sitter}. For each function, we extract the metadata from the corresponding source file, which includes the unique identifier of the function, import libraries, object/class name, input arguments, and output arguments, as shown in Figure \ref{fig:source_code_parse-a}. Then we traverse the AST from the root node to the children nodes corresponding to package-identifier, import-declaration, object-definition, and function-definition nodes, as illustrated in Figure ~\ref{fig:source_code_parse-b}. The functions defined in each source file are located under the children nodes corresponding to the function definitions. We traverse each function-definition node and build a data flow graph for each function based on its variable assignment statements.

\vspace{-10pt}
\subsection{Connecting DFG and Diagnostic Data}

\begin{figure}[t]
\centering{%
   \includegraphics[width=3in]{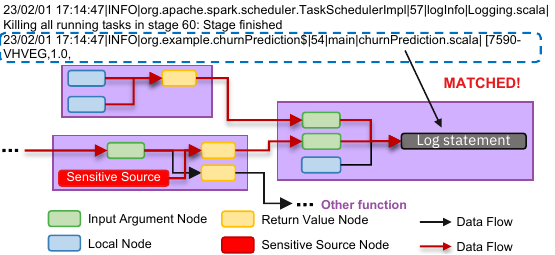}  
}
\vspace{-5pt}
\caption{\label{fig:link-log-dfg}Match the log statement and its corresponding node in the DFG, to enable the backtracking to its data source node.  }
\vspace{-10pt}
\end{figure}

Once the log is generated, to detect the senstiveness of each log message, we need to link the message to the data flow graph by identifying the message's producing log statement. To do so, we use a lightly augmented log message by configuring the logger to output the location of the log statement in the source code. Then, our log analyzer directly query the data flow graph that is corresponding to the function that invoked the log statement and backtrack the ancestors of the function in the graph until it reaches the  data source nodes, as illustrated in Figure ~\ref{fig:link-log-dfg}.

\eat{

}

\section{Evaluation}\label{sec:evaluation}

In this section, we evaluate the effectiveness of our novel redaction framework. Since there are no public benchmark datasets for privacy information redaction. We decided to choose three open-sourced machine learning applications: a telco customer churn prediction \cite{telco-app}, customer segmentation \cite{customer-seg-app}, fraud detection \cite{fraud-det-app}. Sensitive information such as customer profiles and financial information is involved in all three scenarios as well as the diagnostics data (i.e., logs) generated from the three applications. We compare the precision and recall of our approach to three widely-used baseline redaction tools: Google Cloud Data Loss Prevention \cite{google-dlp}, Amazon Macie \cite{amazon-macie}, and Microsoft Presidio \cite{microsoft-presidio}. We run these three applications on Spark and collect logs during the running stage then apply the privacy-preserving log redaction tools. Our framework is implemented in Python with the latest numpy, pandas, and tree\_sitter libraries. As illustrated in Tab.~\ref{tab:precision}, our approach achieves significantly better precision and recall compared to baselines for all three scenarios.

\begin{table}[]
\caption{Precision and Recall Comparison}
\vspace{-5pt}
\label{tab:precision}
\resizebox{\columnwidth}{!}{%
\begin{tabular}{|c||c|c||c|c||c|c|}
\hline
\multirow{2}{*}{Method}           & \multicolumn{2}{c||}{Telcom Churn Prediction} & \multicolumn{2}{c||}{Customer Segmentation} & \multicolumn{2}{c|}{Fraud Detection}  \\ 
\cline{2-7}
                                  & Precision         & Recall                    & Precision         & Recall                  & Precision         & Recall            \\ 
\hline
Google Cloud Data Loss Prevention & 84.73\%  &47.52\%                & 91.96\%  & 10.11\%  & 98.14\%  & 22.13\%      \\ \hline
Amazon Macie                     &        74.12\% & \textbf{100\%}                  &     99.86\%   &     \textbf{98.66\%}                &          99.66\%     &          \textbf{98.81\%}   \\ \hline
Microsoft Presidio                & 12.88\%   & \textbf{100\%}                  & 31.15\%   & 98.65\%             & 6.30\%   & 98.80\%       \\ \hline
Our Approach                              & \textbf{100.00\%}  & \textbf{100\%}                    & \textbf{100.00\%}   & \textbf{98.66\%}            & \textbf{100.00\%} & \textbf{98.81\%}          \\ \hline
\end{tabular}
}\vspace{-10pt}
\end{table}

\eat{
}

\section{Acknowledgment}

This material is based upon work supported by the U.S. Department of Homeland Security under Grant Award Number 17STQAC00001-07-00.

\section{Disclaimer}

The views and conclusions contained in this document are those of the authors and should not be interpreted as necessarily representing the official policies, either expressed or implied, of the U.S. Department of Homeland Security.

\bibliographystyle{ACM-Reference-Format}
\bibliography{refs}


\begin{thebibliography}{19}

%%% the Web address.

\ifx \showCODEN    \undefined \def \showCODEN     #1{\unskip}     \fi
\ifx \showDOI      \undefined \def \showDOI       #1{#1}\fi
\ifx \showISBNx    \undefined \def \showISBNx     #1{\unskip}     \fi
\ifx \showISBNxiii \undefined \def \showISBNxiii  #1{\unskip}     \fi
\ifx \showISSN     \undefined \def \showISSN      #1{\unskip}     \fi
\ifx \showLCCN     \undefined \def \showLCCN      #1{\unskip}     \fi
\ifx \shownote     \undefined \def \shownote      #1{#1}          \fi
\ifx \showarticletitle \undefined \def \showarticletitle #1{#1}   \fi
\ifx \showURL      \undefined \def \showURL       {\relax}        \fi
% invisible to TeX
\providecommand\bibfield[2]{#2}
\providecommand\bibinfo[2]{#2}
\providecommand\natexlab[1]{#1}
\providecommand\showeprint[2][]{arXiv:#2}

\bibitem[\protect\citeauthoryear{??}{ama}{2023}]%
        {amazon-macie}
 \bibinfo{year}{2023}\natexlab{}.
\newblock \bibinfo{title}{Amazon Macie - Sensitive Data Discovery}.
\newblock \bibinfo{howpublished}{\url{https://aws.amazon.com/macie}}.
\newblock


\bibitem[\protect\citeauthoryear{??}{cus}{2023}]%
        {customer-seg-app}
 \bibinfo{year}{2023}\natexlab{}.
\newblock \bibinfo{title}{Customer Segmentation Workloads}.
\newblock \bibinfo{howpublished}{\url{https://www.tpc.org/tpcx-ai}}.
\newblock


\bibitem[\protect\citeauthoryear{??}{fra}{2023}]%
        {fraud-det-app}
 \bibinfo{year}{2023}\natexlab{}.
\newblock \bibinfo{title}{Fraud Detection Workloads}.
\newblock \bibinfo{howpublished}{\url{https://www.tpc.org/tpcx-ai}}.
\newblock


\bibitem[\protect\citeauthoryear{??}{goo}{2023}]%
        {google-dlp}
 \bibinfo{year}{2023}\natexlab{}.
\newblock \bibinfo{title}{Google Cloud Data Loss Prevention}.
\newblock \bibinfo{howpublished}{\url{https://cloud.google.com/dlp}}.
\newblock


\bibitem[\protect\citeauthoryear{??}{mic}{2023}]%
        {microsoft-presidio}
 \bibinfo{year}{2023}\natexlab{}.
\newblock \bibinfo{title}{Microsoft Presidio}.
\newblock \bibinfo{howpublished}{\url{https://github.com/microsoft/presidio}}.
\newblock


\bibitem[\protect\citeauthoryear{??}{ope}{2023}]%
        {open-source-summary}
 \bibinfo{year}{2023}\natexlab{}.
\newblock \bibinfo{title}{A Summary of Census II}.
\newblock
  \bibinfo{howpublished}{\url{https://www.linuxfoundation.org/blog/blog/a-summary-of-census-ii-open-source-software-application-libraries-the-world-depends-on}}.
\newblock


\bibitem[\protect\citeauthoryear{??}{tel}{2023}]%
        {telco-app}
 \bibinfo{year}{2023}\natexlab{}.
\newblock \bibinfo{title}{Telco Customer Churn}.
\newblock
  \bibinfo{howpublished}{\url{https://www.kaggle.com/datasets/blastchar/telco-customer-churn}}.
\newblock


\bibitem[\protect\citeauthoryear{Annas}{Annas}{2003}]%
        {annas2003hipaa}
\bibfield{author}{\bibinfo{person}{George~J Annas}.}
  \bibinfo{year}{2003}\natexlab{}.
\newblock \showarticletitle{HIPAA regulations: a new era of medical-record
  privacy?}
\newblock \bibinfo{journal}{\emph{New England Journal of Medicine}}
  \bibinfo{volume}{348} (\bibinfo{year}{2003}), \bibinfo{pages}{1486}.
\newblock


\bibitem[\protect\citeauthoryear{Chow, Golle, and Staddon}{Chow
  et~al\mbox{.}}{2008}]%
        {chow2008detecting}
\bibfield{author}{\bibinfo{person}{Richard Chow}, \bibinfo{person}{Philippe
  Golle}, {and} \bibinfo{person}{Jessica Staddon}.}
  \bibinfo{year}{2008}\natexlab{}.
\newblock \showarticletitle{Detecting privacy leaks using corpus-based
  association rules}. In \bibinfo{booktitle}{\emph{Proceedings of the 14th ACM
  SIGKDD international conference on Knowledge discovery and data mining}}.
  \bibinfo{pages}{893--901}.
\newblock


\bibitem[\protect\citeauthoryear{Cumby and Ghani}{Cumby and Ghani}{2011}]%
        {cumby2011machine}
\bibfield{author}{\bibinfo{person}{Chad Cumby} {and} \bibinfo{person}{Rayid
  Ghani}.} \bibinfo{year}{2011}\natexlab{}.
\newblock \showarticletitle{A machine learning based system for
  semi-automatically redacting documents}. In
  \bibinfo{booktitle}{\emph{Proceedings of the AAAI Conference on Artificial
  Intelligence}}, Vol.~\bibinfo{volume}{25}. \bibinfo{pages}{1628--1635}.
\newblock


\bibitem[\protect\citeauthoryear{He, Zhu, He, and Lyu}{He
  et~al\mbox{.}}{2020}]%
        {he2020loghub}
\bibfield{author}{\bibinfo{person}{Shilin He}, \bibinfo{person}{Jieming Zhu},
  \bibinfo{person}{Pinjia He}, {and} \bibinfo{person}{Michael~R Lyu}.}
  \bibinfo{year}{2020}\natexlab{}.
\newblock \showarticletitle{Loghub: a large collection of system log datasets
  towards automated log analytics}.
\newblock \bibinfo{journal}{\emph{arXiv preprint arXiv:2008.06448}}
  (\bibinfo{year}{2020}).
\newblock


\bibitem[\protect\citeauthoryear{Jones, Kumar, Pang, and Tomkins}{Jones
  et~al\mbox{.}}{2007}]%
        {jones2007know}
\bibfield{author}{\bibinfo{person}{Rosie Jones}, \bibinfo{person}{Ravi Kumar},
  \bibinfo{person}{Bo Pang}, {and} \bibinfo{person}{Andrew Tomkins}.}
  \bibinfo{year}{2007}\natexlab{}.
\newblock \showarticletitle{" I know what you did last summer" query logs and
  user privacy}. In \bibinfo{booktitle}{\emph{Proceedings of the sixteenth ACM
  conference on Conference on information and knowledge management}}.
  \bibinfo{pages}{909--914}.
\newblock


\bibitem[\protect\citeauthoryear{Kaul, Kesarwani, Min, and Zhang}{Kaul
  et~al\mbox{.}}{2021}]%
        {kaul2021knowledge}
\bibfield{author}{\bibinfo{person}{Akshar Kaul}, \bibinfo{person}{Manish
  Kesarwani}, \bibinfo{person}{Hong Min}, {and} \bibinfo{person}{Qi Zhang}.}
  \bibinfo{year}{2021}\natexlab{}.
\newblock \showarticletitle{Knowledge \& learning-based adaptable system for
  sensitive information identification and handling}. In
  \bibinfo{booktitle}{\emph{2021 IEEE 14th International Conference on Cloud
  Computing (CLOUD)}}. IEEE, \bibinfo{pages}{261--271}.
\newblock


\bibitem[\protect\citeauthoryear{Mouza, M{\'e}tais, Lammari, Akoka, Aubonnet,
  Comyn-Wattiau, Fadili, and Cherfi}{Mouza et~al\mbox{.}}{2010}]%
        {mouza2010towards}
\bibfield{author}{\bibinfo{person}{C{\'e}dric Mouza},
  \bibinfo{person}{Elisabeth M{\'e}tais}, \bibinfo{person}{Nadira Lammari},
  \bibinfo{person}{Jacky Akoka}, \bibinfo{person}{Tatiana Aubonnet},
  \bibinfo{person}{Isabelle Comyn-Wattiau}, \bibinfo{person}{Hammou Fadili},
  {and} \bibinfo{person}{Samira Si-Sa{\"\i}d Cherfi}.}
  \bibinfo{year}{2010}\natexlab{}.
\newblock \showarticletitle{Towards an automatic detection of sensitive
  information in a database}. In \bibinfo{booktitle}{\emph{2010 Second
  International Conference on Advances in Databases, Knowledge, and Data
  Applications}}. IEEE, \bibinfo{pages}{247--252}.
\newblock


\bibitem[\protect\citeauthoryear{Park, Gates, Teiken, and Cheng}{Park
  et~al\mbox{.}}{2011}]%
        {park2011experimental}
\bibfield{author}{\bibinfo{person}{Youngja Park}, \bibinfo{person}{Stephen~C
  Gates}, \bibinfo{person}{Wilfried Teiken}, {and} \bibinfo{person}{Pau-Chen
  Cheng}.} \bibinfo{year}{2011}\natexlab{}.
\newblock \showarticletitle{An experimental study on the measurement of data
  sensitivity}. In \bibinfo{booktitle}{\emph{Proceedings of the first workshop
  on building analysis datasets and gathering experience returns for
  security}}. \bibinfo{pages}{70--77}.
\newblock


\bibitem[\protect\citeauthoryear{Regulation}{Regulation}{2018}]%
        {regulation2018general}
\bibfield{author}{\bibinfo{person}{Protection Regulation}.}
  \bibinfo{year}{2018}\natexlab{}.
\newblock \showarticletitle{General data protection regulation}.
\newblock \bibinfo{journal}{\emph{Intouch}}  \bibinfo{volume}{25}
  (\bibinfo{year}{2018}), \bibinfo{pages}{1--5}.
\newblock


\bibitem[\protect\citeauthoryear{S{\'a}nchez and Batet}{S{\'a}nchez and
  Batet}{2016}]%
        {sanchez2016c}
\bibfield{author}{\bibinfo{person}{David S{\'a}nchez} {and}
  \bibinfo{person}{Montserrat Batet}.} \bibinfo{year}{2016}\natexlab{}.
\newblock \showarticletitle{C-sanitized: A privacy model for document redaction
  and sanitization}.
\newblock \bibinfo{journal}{\emph{Journal of the Association for Information
  Science and Technology}} \bibinfo{volume}{67}, \bibinfo{number}{1}
  (\bibinfo{year}{2016}), \bibinfo{pages}{148--163}.
\newblock


\bibitem[\protect\citeauthoryear{Virtue}{Virtue}{2009}]%
        {virtue2009payment}
\bibfield{author}{\bibinfo{person}{Timothy~M Virtue}.}
  \bibinfo{year}{2009}\natexlab{}.
\newblock \bibinfo{booktitle}{\emph{Payment card industry data security
  standard handbook}}.
\newblock \bibinfo{publisher}{Wiley Online Library}.
\newblock


\bibitem[\protect\citeauthoryear{Wang, Ko, Zhang, Yang, and Lin}{Wang
  et~al\mbox{.}}{2023}]%
        {chao2023flow}
\bibfield{author}{\bibinfo{person}{Chao Wang}, \bibinfo{person}{Ronny Ko},
  \bibinfo{person}{Yue Zhang}, \bibinfo{person}{Yuqing Yang}, {and}
  \bibinfo{person}{Zhiqiang Lin}.} \bibinfo{year}{2023}\natexlab{}.
\newblock \showarticletitle{TAINTMINI: Detecting Flow of Sensitive Data in
  Mini-Programs with Static Taint Analysis}. In
  \bibinfo{booktitle}{\emph{ICSE}}.
\newblock


\end{thebibliography}

\end{document}